\documentclass[twocolumn]{pasj00}
\SetRunningHead{Astronomical Society of Japan}{Usage of \texttt{pasj00.cls}}

\usepackage{times}

\begin{document}

\title{A Study of the Long-term Spectral Variations of 3C 66A \\Observed
  with the \emph{Fermi} and Kanata Telescopes}

\author{
Ryosuke~\textsc{Itoh}\altaffilmark{1,2}, 
Yasushi~\textsc{Fukazawa}\altaffilmark{1,3}, 
James~\textsc{Chiang}\altaffilmark{4}, 
Yoshitaka~\textsc{Hanabata}\altaffilmark{1}, 
Masaaki~\textsc{Hayashida}\altaffilmark{4,5}, 
Katsuhiro~\textsc{Hayashi}\altaffilmark{1},
Tsunefumi~\textsc{Mizuno}\altaffilmark{6}, 
Masanori~\textsc{Ohno}\altaffilmark{1}, 
Takashi~\textsc{Ohsugi}\altaffilmark{1}, 
Jeremy~S.~\textsc{Perkins}\altaffilmark{7,8,9,10}, 
Silvia~\textsc{Rain\`o}\altaffilmark{11,12}, 
Luis~C.~\textsc{Reyes}\altaffilmark{13}, 
Hiromitsu~\textsc{Takahashi}\altaffilmark{1}, 
Yasuyuki~\textsc{Tanaka}\altaffilmark{6}, 
Gino~\textsc{Tosti}\altaffilmark{14,15},

Hiroshi~\textsc{Akitaya}\altaffilmark{6}, 
Akira~\textsc{Arai}\altaffilmark{16}, 
Masaru~\textsc{Kino}\altaffilmark{17}, 
Yuki~\textsc{Ikejiri}\altaffilmark{1}, 
Koji~S.~\textsc{Kawabata}\altaffilmark{6}, 
Tomoyuki~\textsc{Komatsu}\altaffilmark{1}, 
Kiyoshi~\textsc{Sakimoto}\altaffilmark{1}, 
Mahito~\textsc{Sasada}\altaffilmark{18}, 
Shuji~\textsc{Sato}\altaffilmark{17}, 
Makoto~\textsc{Uemura}\altaffilmark{6}, 
Takahiro~\textsc{Ui}\altaffilmark{1}, 
Masayuki~\textsc{Yamanaka}\altaffilmark{19} 
and
Michitoshi~\textsc{Yoshida}\altaffilmark{6}
}

\altaffiltext{1}{Department of Physical Sciences, Hiroshima University, Higashi-Hiroshima, Hiroshima 739-8526, Japan}
\altaffiltext{2}{email: itoh@hep01.hepl.hiroshima-u.ac.jp}
\altaffiltext{3}{email: fukazawa@hep01.hepl.hiroshima-u.ac.jp}
\altaffiltext{4}{W. W. Hansen Experimental Physics Laboratory, Kavli Institute for Particle Astrophysics and Cosmology, Department of Physics and SLAC National Accelerator Laboratory, Stanford University, Stanford, CA 94305, USA}
\altaffiltext{5}{Department of Astronomy, Graduate School of Science, Kyoto University, Sakyo-ku, Kyoto 606-8502, Japan}
\altaffiltext{6}{Hiroshima Astrophysical Science Center, Hiroshima University, Higashi-Hiroshima, Hiroshima 739-8526, Japan}
\altaffiltext{7}{NASA Goddard Space Flight Center, Greenbelt, MD 20771, USA}
\altaffiltext{8}{Department of Physics and Center for Space Sciences and Technology, University of Maryland Baltimore County, Baltimore, MD 21250, USA}
\altaffiltext{9}{Center for Research and Exploration in Space Science and Technology (CRESST) and NASA Goddard Space Flight Center, Greenbelt, MD 20771, USA}
\altaffiltext{10}{Harvard-Smithsonian Center for Astrophysics, Cambridge, MA 02138, USA}
\altaffiltext{11}{Dipartimento di Fisica ``M. Merlin" dell'Universit\`a e del Politecnico di Bari, I-70126 Bari, Italy}
\altaffiltext{12}{Istituto Nazionale di Fisica Nucleare, Sezione di Bari, 70126 Bari, Italy}
\altaffiltext{13}{Department of Physics, California Polytechnic State University, San Luis Obispo, CA 93401, USA}
\altaffiltext{14}{Istituto Nazionale di Fisica Nucleare, Sezione di Perugia, I-06123 Perugia, Italy}
\altaffiltext{15}{Dipartimento di Fisica, Universit\`a degli Studi di Perugia, I-06123 Perugia, Italy}
\altaffiltext{16}{Department of Physics, Kyoto Sangyo University, Kyoto, Japan}
\altaffiltext{17}{Department of Physics and Astrophysics, Nagoya University, Chikusa-ku Nagoya 464-8602, Japan}
\altaffiltext{18}{Department of Physics, Graduate School of Science, Kyoto University, Kyoto, Japan}
\altaffiltext{19}{Kwasan Observatory, Kyoto University, Kyoto, Japan}


\maketitle

\begin{abstract}

 3C 66A is an intermediate-frequency-peaked BL Lac object detected by
 the Large Area Telescope onboard the \emph{Fermi Gamma-ray Space
 Telescope}. 
 We present a study of the long-term 
 variations of this blazar seen over $\sim$2 years at GeV energies
 with {\it Fermi} and in the
 optical (flux and polarization) and near infrared with the Kanata
 telescope.  In 2008, the first year of the study, 
 we find a correlation
 between the gamma-ray flux and the measurements taken with the
 Kanata telescope.  This is in contrast to the later measurements
 performed during 2009--2010 which show only a weak correlation along
 with a gradual increase of the optical flux. We calculate an external
 seed photon energy density assuming that the gamma-ray emission is
 due to external Compton scattering.  The energy density of the
 external photons is found to be higher by a factor of two in 2008
 compared to 2009--2010.  We conclude that the different behaviors
 observed between the first year and the later years might be
 explained by postulating two different emission components.

\end{abstract}

\section{Introduction}

Blazars are highly variable active galactic nuclei (AGN) detected at
all wavelengths from radio to gamma rays.  They have strong
relativistic jets aligned with the observer's line of sight and are
apparently bright due to relativistic beaming . Their emission
typically consists of two spectral components. The first, highly
polarized, one is attributed to synchrotron radiation from
relativistic electrons and is emitted at lower energies (typically
radio to UV).  The other, higher energy, component (typically X-ray to
TeV), is not fully understood.  The most plausible scenario is that
the emission is due to inverse Compton 
scattering off some combination of synchrotron and external photons.

3C 66A is one of the most famous TeV blazars, and is classified as an
intermediate-frequency-peaked BL Lac object
(IBL; \cite{Schlegel98,Abdo10a}).  Its low-energy component extends
from radio to soft X-rays with a peak in the optical band
\citep{Abdo10a}.  Most TeV BL Lac objects are classified as HBL
(high-frequency-peaked BL Lac), and therefore 3C 66A, as an IBL, is a
rare and valuable source which can be used to study the emission
mechanisms of high-energy gamma-rays. The high-energy component
located in the MeV to TeV gamma-ray band, can be explained by a
Synchrotron Self-Compton (SSC) plus External Compton (EC) model
\citep[and references therein, referred to as ``Paper I'']{Abdo10d}.
3C 66A is listed as 2FGL J0222.6+4302 in the 
\emph{Fermi} Large Area Telescope (LAT) second source catalog \citep{Abdo12Cat}. 
Due to a lack of strong emission lines in
the optical band, the redshift is uncertain. \citet{Miller78} reported
a redshift {\it z} = 0.444 based on a detection of the Mg\emissiontype{II}
line.  However, as pointed out by \citet{Bramel05}, this redshift is
unreliable and should be reported as {\it z} = 0.11$\sim$0.444 because
it was determined by only a single weak line.  The redshift
uncertainty prevents a definite correction of the TeV gamma-ray
absorption by the Extragalactic Background Light (EBL).  Thus, the
true spectrum is not well known and the emission mechanism in the TeV
gamma-ray band is not yet understood.  Thus, multiwavelength
observations are an extremely powerful tool to probe the emission
mechanism.

The launch of the \emph{Fermi Gamma-ray Space Telescope} started a new
era of multiwavelength monitoring observations of AGN.  In
particular, the all-sky monitoring capability of the \emph{Fermi}-LAT
Telescope has prompted extensive simultaneous gamma-ray and optical
observing campaigns. These are important since correlation studies
between the GeV gamma-ray and optical bands of IBLs provide
information on the source electron population. Optical polarization
measurements are particularly important since they can probe jet
structures.  In this paper, we report a correlation study of the GeV
gamma-ray and optical emission of 3C 66A, based on simultaneous
observations by the Large Area Telescope onboard the
\emph{Fermi Gamma-ray Space Telescope} and the Kanata optical
telescope.  We assume a flat $\Lambda$CDM cosmology with $H_{0}=71$ km
s$^{-1}$Mpc$^{-1}$ and $\Omega_m=0.27$ \citep{Wright06}.

\section{Observations and Results}
\subsection{Fermi observations and Data analysis} \label{sec:fermi}
\begin{figure}[!htb]
\begin{center}
\FigureFile(80mm,80mm){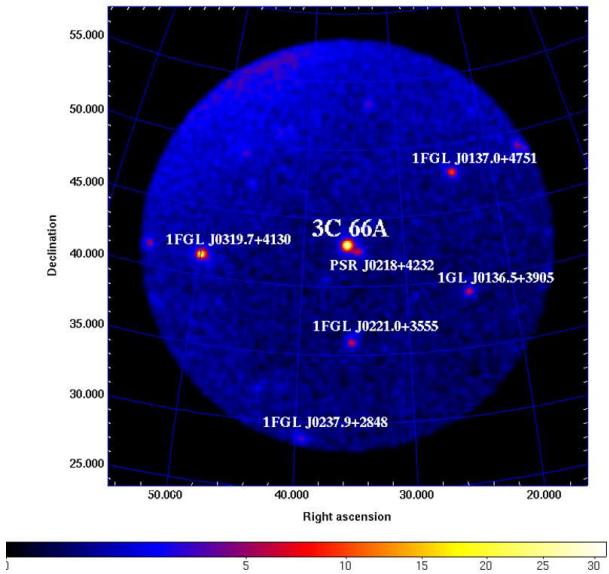}
\end{center}
\caption{Smoothed count map of the 3C 66A region as seen by \emph{Fermi}-LAT between 2008 August and 2010 February with data above 1 GeV.} 
\label{fig:CMAP}
\end{figure}

The LAT is the main instrument
onboard the \emph{Fermi Gamma-ray Space Telescope}.  The LAT is an
electron-positron pair production detector which is sensitive to
gamma rays with energies from 20 MeV to $>$ 300
GeV.  The LAT observes the whole sky every 3 hours with a large
effective area of $8000$ cm$^2$ at 1 GeV, a wide field of view of
$2.4$ sr, and a single photon angular resolution (68\% containment radius) of 0.1$\arcdeg$ at 10 GeV.
Details of the LAT instrument are described in \citet{Atwood09}.
The data used in this analysis were taken between 2008
August and 2010 February almost entirely in sky survey mode.  
The data were analyzed using the standard \emph{Fermi} analysis software
(\verb|Science Tools|, version v9r15, IRFs P6$\_$V3
 \footnote{http://fermi.gsfc.nasa.gov/ssc/data/analysis/documentation/Cicerone/\\Cicerone\_LAT\_IRFs/IRF\_overview.html}
 ). In order to limit residual contamination of charged-particle
 backgrounds, only ``Diffuse'' class event data above 100 MeV were
 selected. ``Diffuse'' class events are highly likely to be
 photons. 
We also restricted our analysis to events with zenith angles
 $<$105$\arcdeg$ and rocking angles of the LAT $<$52$\arcdeg$ 
to limit the contamination by gamma rays from the Earth's limb.
We performed an {\it Unbinned Likelihood} analysis to calculate the
 gamma-ray spectrum and flux of 3C 66A, using the \verb|gtlike|
 package in the \verb|Science Tools|.  An area of 15$\arcdeg$ around 3C
 66A was selected as a region of interest (ROI) for this analysis. We
 constructed a model of the ROI that includes a point source at the
 position of 3C 66A (R.A. = 35.662$\arcdeg$, Dec. = 43.036$\arcdeg$, J2000)
 , the Galactic diffuse emission component ({\tt gll\_iem\_v02.fit})
 and the isotropic diffuse emission component ({\tt
   isotropic\_iem\_v02.txt}, this is the sum of the extragalactic
 diffuse gamma-ray and the residual charged particle background).
 There are also six bright (Test Statistic $>$ 100)\footnote{The Test
   Statistic (TS) is defined as TS $= -2(\log L0 -\log L1)$ with $L0$
   the likelihood of the Null hypothesis model as compared to the
   likelihood of a competitive model $L1$} point sources within
 15$\arcdeg$ of 3C 66A which were included in the model, as shown in
 Figure \ref{fig:CMAP} and Table \ref{tab:source}. 
The other sources with a TS values of $<$
 100 in \emph{Fermi} LAT first source catalog \citep{Abdo10Cat} are not included in the model.  
 It is necessary to account for
 the emission from the pulsar PSR J0218+4232, located at 0.96$\arcdeg$ from
 3C 66A.  Due to the small point spread function of the LAT
 \citep{Atwood09}, we can separate the emission from the pulsar from
 the emission from 3C 66A and obtain an accurate measurement of the
 GeV gamma-ray flux of 3C 66A without a risk of contamination from the
 pulsar.  The gamma-ray emission from PSR J0218+4232, can be
 represented by the following pulsar super exponential cutoff ({\tt
   PLSuperExpCutoff}) function.
\begin{equation}
 \frac{dN}{dE} =  N_0 \left( \frac{E}{E_0} \right)^{-\gamma} \exp\left( -
							 \frac{E}{E_c}\right) 
\end{equation}
We fixed $\gamma$, $E_0$, and $E_c$ to the values found in
\citet{Abdo10c}: $\gamma = 2.02$, $E_0=1$ GeV and $E_c=5.1$ GeV
.

We verified that the flux of PSR J0218+4232 is stable by creating a
light curve with 50 day time bins, and confirmed that the flux is
statistically constant within the errors.  A $\chi^2$ test for a
constant fit to the light curve of PSR J0218+4232 yielded a
$\chi^2$/d.o.f = 4.008/7.  Hereafter, we fixed the model parameters of
the spectral shape of PSR J0218+4232 to those listed above.

\begin{table}[!htb]
 \caption{Background point sources included in the model.}
\label{tab:source}
\ \\
\centering
 {\small 
   \begin{tabular}{cc}\hline\hline
    Source name & Separation from 3C~66A [deg] \\ \hline
    1FGL J0136.5+3905 & 9.5  \\
    1FGL J0137.0+4751 & 9.3  \\
    1FGL J0221.0+3555 & 7.1  \\
    1FGL J0237.9+2848 & 14.6 \\
    1FGL J0319.7+4130 & 10.6 \\
    PSR J0218+4232    & 0.96 \\ 
    \hline
   \end{tabular}
 }
\end{table}

We modeled the spectrum of 3C 66A as a power-law:
\begin{equation}
\frac{dN}{dE} = N_0 \left(\frac{E}{E_0}\right)^{-\gamma}  \label{eq:pl}
\end{equation}
where $N_0$ is the normalization at energy $E_0 = 100$MeV and $\gamma$ is the
photon index.  The same model was used for the other point sources in
the ROI, except for PSR J0218+4232.  The integrated photon flux
between 100 MeV and 100 GeV is calculated by integrating equation
\ref{eq:pl}. 
In the source model for the likelihood analysis, the spectral indicies of
the background sources were fixed to their catalog values while their
normalizations were left free.  

\begin{figure*}[!htb]
\begin{center}
\FigureFile(150mm,50mm){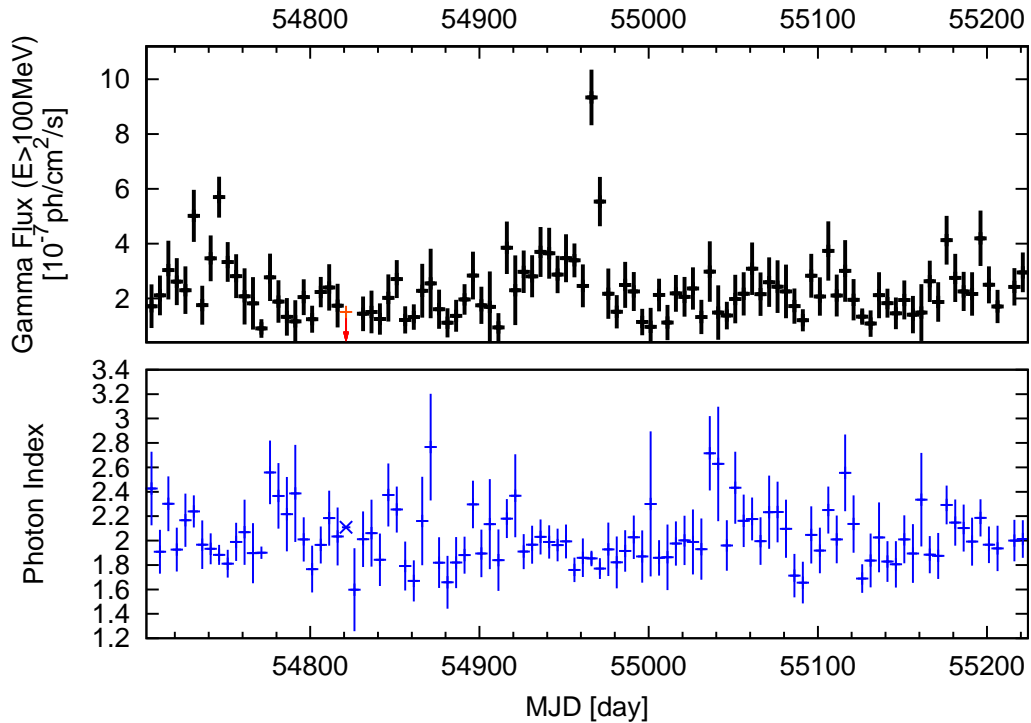}
\end{center}
\caption{GeV Gamma-ray light curve of 3C 66A measured by the
  \emph{Fermi} LAT from MJD 54683 to 55232 (2008-08-04 to
  2010-02-04). One bin corresponds to 5 days. The upper
    panel shows the integral photon flux from
    100MeV--100GeV, and the lower panel shows the photon index.
    Bins with TS values smaller than
    10 are indicated by arrows and are 90\% C.L. upper limits.}
\label{fig:LC}
\end{figure*}

\begin{figure}[!htb]
\begin{center}
\FigureFile(80mm,50mm){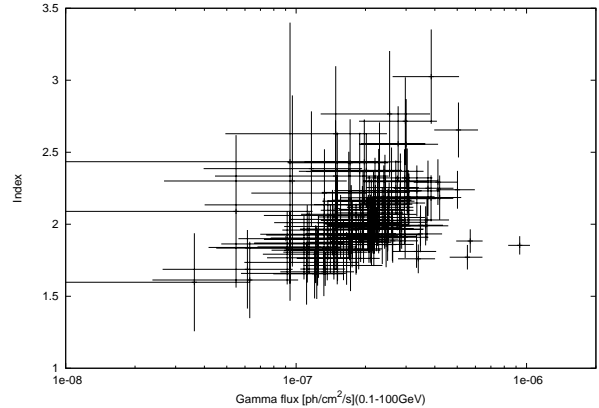}
\end{center}
\caption{GeV gamma-ray flux vs. photon index.}
 \label{fig:GvsI}
\end{figure}

Figure \ref{fig:LC} shows how the flux and photon index of 3C 66A
varies with time in the 100 MeV to 100 GeV energy band calculated by
performing a likelihood analysis in 5-day time intervals.  
Data points which have small TS values ($<$ 10) are designated by arrows and
indicate 90\% C.L. upper limits. Two flares are evident in the light
curve, one around MJD 54750 (hereafter ``flare 1''), and the other one
around MJD 54960 (``flare 2'').  Flare 2 is not considered in this
paper because simultaneous optical data are not available due to the
proximity of 3C 66A to the Sun during the flare.  The maximum flux
during flare 1, which is also discussed in Paper I, is $5 \times
10^{-7}$ ph cm$^{-2}$s$^{-1}$ from 0.1--100 GeV.  Except during these
flares, the flux of 3C 66A is around $\sim 2 \times 10^{-7}$ ph
cm$^{-2}$ s$^{-1}$ (0.1--100 GeV), as reported in Paper I.  On the
other hand, the photon index does not change significantly (a $\chi^2$
test with a constant fit gives a $\chi^2/$d.o.f = 138.5/121) and no
clear correlation with the flux is seen. Figure \ref{fig:GvsI}
illustrates a trend; the spectrum is softer when the flux is
higher. 
However, this could be parameter coupling since there is a mathematical 
correlation between the integral flux and index; the integral flux is 
obtained from integrating equation 2.   
In addition, the sensitivity of the LAT is greater for harder spectrum objects.
This correlation can also be caused by observational bias,
so we cannot conclude that this correlation is significant.
This is consistent with the findings of  \citet{Abdo10i}, who state
that intermediate synchrotron peaked blazars (ISPs) in general show
only a small change of photon index ($<$ 0.2) with flux variations.

\subsection{Kanata Telescope Observations and Optical properties}\label{sec:kanata}
We performed {\it V}, {\it J}, {\it Ks}-band photometry and polarimetry
observations of 3C 66A from July 2008 to February 2010, using the
TRISPEC instrument installed on the 1.5m Kanata telescope located at
the Higashi-Hiroshima Observatory in Japan.  The TRISPEC consists of a
CCD and two InSb arrays, enabling photopolarimetric observations in
one optical and two NIR bands simultaneously \citep{Watanabe05}.

We obtained 278 photometry and 208 polarimetry observations in the
{\it V} band, 180 photometry observations in the {\it J} band, and 77
photometry observations in the {\it Ks} band.  Each observing
sequence consisted of successive exposures at four position angles of
a half-wave plate; $0\arcdeg, 45\arcdeg, 22.5\arcdeg, $and
$67.5\arcdeg$.  The data reduction involved standard CCD photometry
procedures; aperture photometry using \verb|APPHOT| packaged in
\verb|PYRAF| 
\footnote{PYRAF is a product of the Space Telescope Science Institute,
which is operated by AURA for NASA.}
\footnote{ http://www.stsci.edu/institute/software\_hardware/pyraf}
, and differential photometry with a comparison star taken
in the same frame of 3C 66A.  
The position of the comparison star is R.A. $=02^h22^m55.12^s$ and
Dec$=+43\arcdeg 03\arcmin 15.5 \arcsec$ (J2000), and its magnitudes
are {\it V} = 12.809, {\it J} = 12.371, {\it Ks} = 12.282
(\cite{Gonzalez01}, \cite{Cutri03}).  The data have been corrected for
Galactic extinction using $A({\it V})$=0.274 and $A({\it J})$=0.076.
We confirmed that the instrumental polarization was smaller than 0.1\%
in the V band, using unpolarized standard stars and thus applied no
correction for it. The polarization angle (PA) is defined in the
standard manner (measured from north to east), by calibrations with
polarized stars, HD19820 and HD25443 \citep{Wolff96}.  Because the PA
has an ambiguity of $\pm 180\arcdeg \times n$ (where $n$ is an
integer), we selected $n$ which gives the least angle difference from
previous measurements, assuming that the PA would change smoothly.

\begin{figure*}[!htb]
\begin{center}
\FigureFile(150mm,50mm){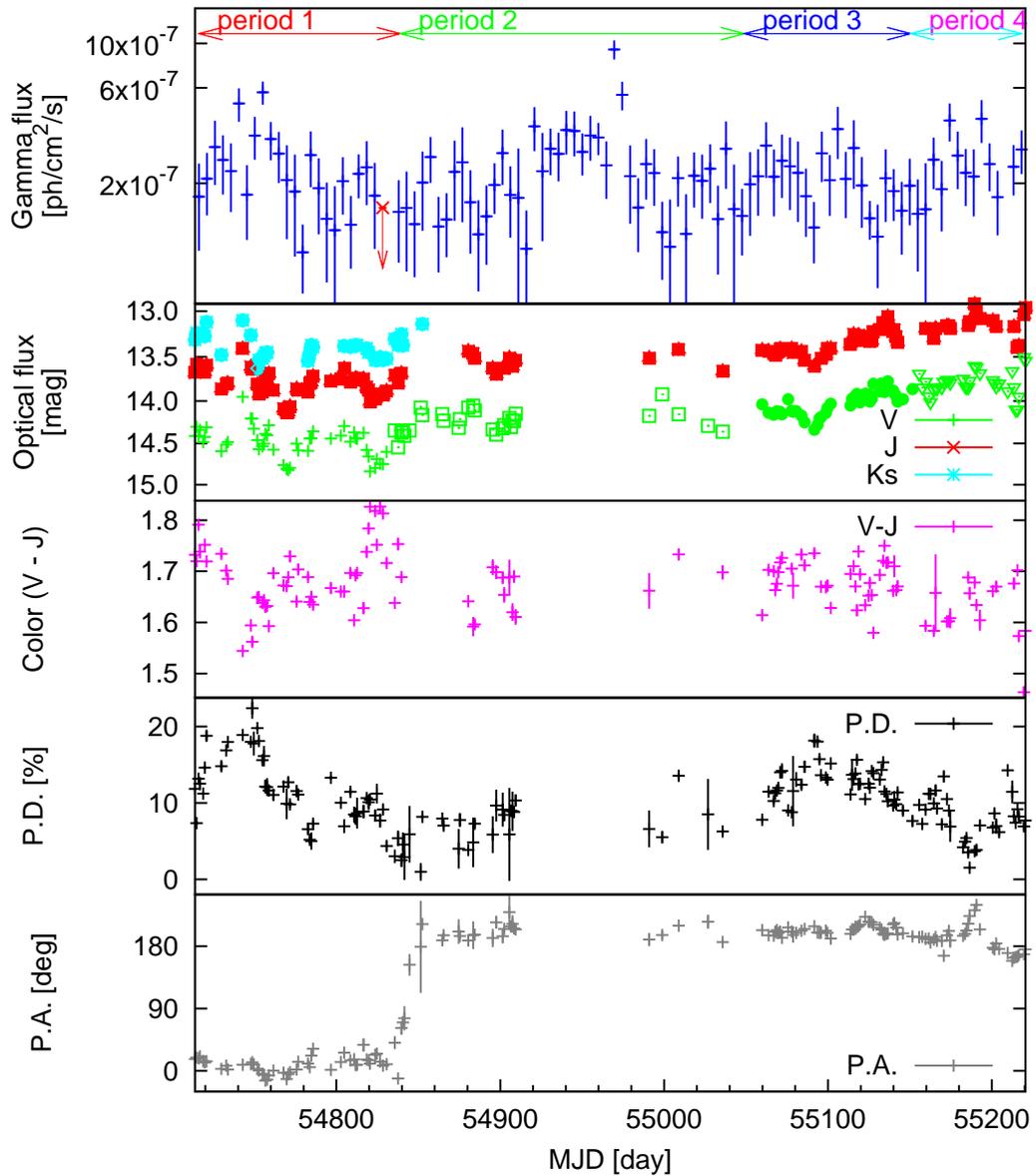}
\end{center}
\caption{ Optical flux and polarization of 3C 66A plotted versus time
  as measured by the Kanata telescope.  The Gamma-ray light curve
  (100MeV--100GeV) is shown in the top panel for comparison.  Bins
  with TS values smaller than 10 are indicated by arrows and are 90\%
  C.L. upper limits. The second panel is the optical {\it V,
    J}and,{\it Ks}-band flux in units of magnitude. The third panel
  shows the optical color index ({\it V} band - {\it J} band). The
  fourth panel is the {\it V}-band polarization degree (PD) and the
  bottom panel is the polarization angle (PA).}
 \label{fig:LCopt}
\end{figure*}

Figure \ref{fig:LCopt} shows how the optical and NIR flux and
polarization change with time. The GeV gamma-ray light curve from
Figure \ref{fig:LC} is reproduced in panel 1 for comparison.  Several
polarization and flux states are seen in the optical data.  Based on
the optical flux and polarization degree (PD), we define four periods; MJD
54705--54830 (hereafter ``period 1''), MJD 54831--55047 (``period
2''), MJD 55048--55150 (``period 3'') and MJD 55151--55220 (``period
4''). 
We collect the properties of variability for each period.
In period 1, both the gamma-ray flux at 20 GeV and the optical
{\it V} band flux ($F_{G1},F_{V1}$) are low at an average of 
$F_{G1} = (3.5 \pm 1.6) \times 10^{-11}$ erg cm$^{-2}$s$^{-1}$, 
$F_{V1} = (3.2 \pm 0.9) \times 10^{-11}$ erg cm$^{-2}$s$^{-1}$ 
respectively, with the exception of the flare 1.
During period 1 the optical properties vary violently and the gamma-ray 
flux is well correlated with the optical flux, color, and polarization 
degree.

\begin{figure*}[!htb]
\begin{center}
\FigureFile(150mm,50mm){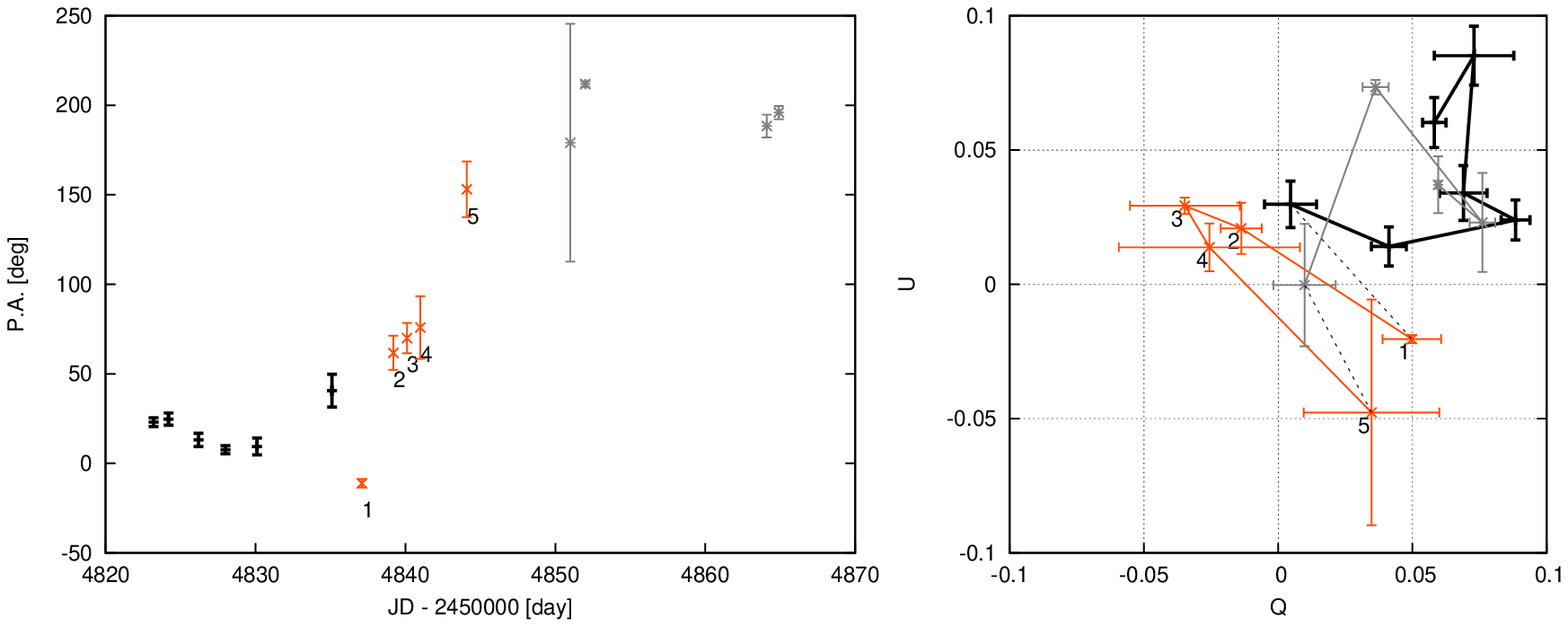}
\end{center}
\caption{The left plot shows the PA versus time for the period around
  MJD 54830 and the right plot shows the QU-plane for this same time
  period.  The plots are color and number coded so that specific time
  periods can be referenced.}
\label{fig:QU}
\end{figure*}

On MJD 54830 (the beginning of period 2), the optical flux suddenly
increased and a high-optical-flux state began. The optical coverage
during parts of period 2 are incomplete because 3C 66A was located
near the Sun.  Thus correlation studies during this time are
problematic.  The polarization degree became relatively low (its
average value in periods 1 and 2 are $PD_{1} = 11.9$ $\%$ and $PD_{2}
= 6.6$ \%, respectively) and the PA is also constant.  From MJD 54835
to MJD 54850, the PA changed by almost $180\arcdeg$ in 2 weeks.  Such
a rotation of PA is similar to that found for BL~Lac and 3C~279
\citep{Marscher08,Abdo10e}.  Figure \ref{fig:QU} shows the variation
of PA and Stokes parameters $Q$ and $U$ from MJD 54835 to MJD 54850.
The variation on the QU plane implies rotation, but it could be due to
random motion around the origin of the QU plane.  Because of the
aforementioned observation sparseness, we cannot discuss this behavior
further.

In period 3, continuous optical observations are available, and the PD
became relatively high (the average value in period 3 is $PD_{3} =
12.4$ \%).

In period 4, the PD became lower (the average value is $PD_{4} = 8.0$
\%) and the color reddened.  The optical flux was higher than in the
other periods but the gamma-ray flux ($F_{G4} = (3.8 \pm 1.6) \times
10^{-11}$ erg cm$^{-2}$s$^{-1}$) was similar to that in other periods.
A weak flare in the optical band was seen on MJD 55190, associated
with a decrease of the PD and a temporary shift of the PA by $\sim$40$\arcdeg$.

\subsection{Correlation Studies} \label{sec:correlation} 

Correlations between the various gamma-ray and optical properties can
be quantified by calculating the discrete correlation function (DCF)
\citep{Edelson98}. Since the gamma-ray flux is averaged over 5-day
intervals, it was necessary to also average the optical data.  The
significance of the correlation (95\% C.L.) was tested using the
Student's t-test. A test statistic for the t-test ($TS_t$) was
calculated as follows,
\begin{equation}
TS_{t} = \frac{DCF}{\sqrt{1-DCF^2}}\times\sqrt{n-2}
\end{equation}
where $DCF$ is the DCF value and $n$ is the number of degrees of
freedom.  The DCF value for zero time lag corresponds to the
correlation coefficient.  If the $TS_t$ value is larger than the
critical values of the t-distribution for a significance level of
$\alpha$ = 0.95 against the null hypothesis and the DCF value is
larger than 0.5, then the correlation is considered significant.

\begin{figure}[!htb]
\begin{center}
\FigureFile(80mm,50mm){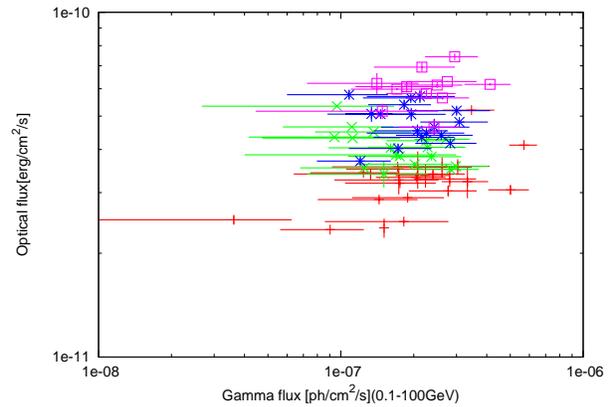}
\end{center}
\caption{Gamma-ray flux vs. Optical {\it V} band flux. Colors
 indicate four periods; red: period 1 (MJD 54706--54831), green: period 2.(MJD
 54831--55047), blue: period 3 (MJD 55047--55151), and pink: 
period 4 (MJD 55151--55220).}
 \label{fig:GvsV}
\end{figure}

\begin{figure}[!htb]
\begin{center}
\FigureFile(80mm,50mm){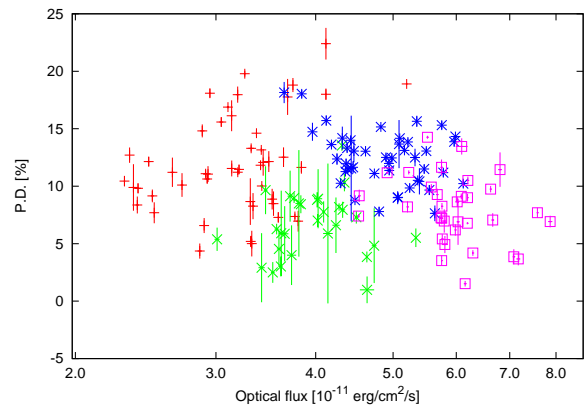}
\end{center}
\caption{Optical {\it V} band flux vs. optical polarization degree (PD).
Colors are the same as those of Fig.\ref{fig:GvsV}.}  
\label{fig:VvsPD}
\end{figure}

\begin{figure}[!htb]
\begin{center}
\FigureFile(80mm,50mm){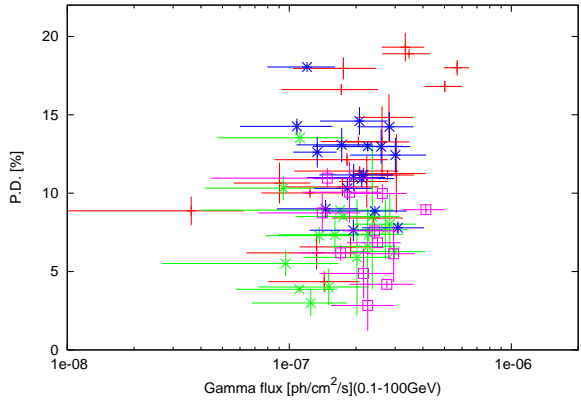}
\end{center}
\caption{Gamma-ray flux vs. optical polarization degree (PD).
Colors are the same as those of Fig.\ref{fig:GvsV}.}
 \label{fig:GvsPD}
\end{figure}

\begin{figure}[!htb]
\begin{center}
\FigureFile(80mm,50mm){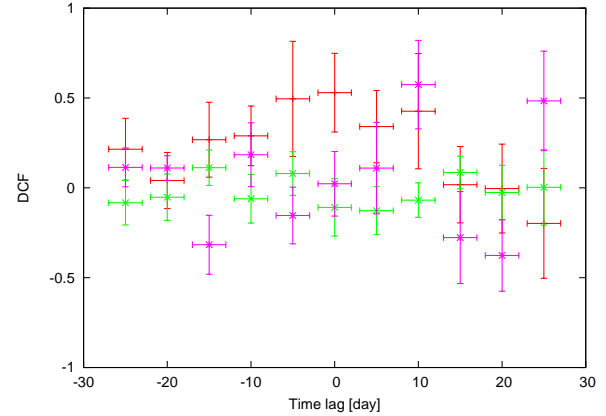}
\FigureFile(80mm,50mm){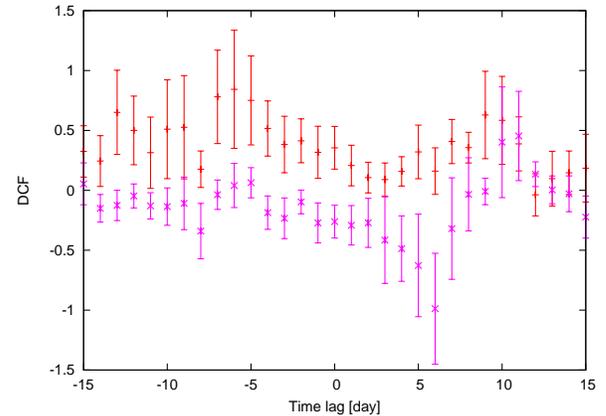}
\end{center}
\caption{Discrete correlation function (DCF) between the gamma-ray
  flux and optical flux (left plot), and the
  optical flux and PD (right plot). Colors indicate four
  periods; red: period 1 (MJD 54706--54831), green: period 2.(MJD
  54831--55047), blue: period 3 (MJD 55047--55151), and pink: period 4
  (MJD 55151--55220). The DCF in periods 2, 3 in Figure (b) are not shown  because of large
  errors.  The horizontal axis represents the time
  lag.}
 \label{fig:DCF}
\end{figure}

In Table \ref{tab:correlation}, we summarize the correlation
coefficients and t-test results for each combination of observed
values.  Only period 1 shows a possible correlation between the
gamma-ray and optical flux.  This correlation can be seen in
Fig. \ref{fig:GvsV} which plots the gamma-ray flux versus the optical
flux and designates each period with different colors. The correlation
in period 1 reported here is also suggested in Fig. \ref{fig:GvsV} and
is consistent with previous measurements (Paper I).  The correlation
between the optical flux and optical polarization degree at zero lag is not
significant in any of the periods.  Fig. \ref{fig:DCF} shows the DCF
of the PD against the optical flux for each period.  The PD variation
in period 1 seems to correlate with the optical flux with some time
offsets; the extrema on MJD 54750, 54770, 54830, and 55190, are seen
in both the optical flux and PD.  It can be seen that the PD
correlated with the optical flux with a 5--7 day delay in period 1 and
anti-correlated with a $\sim$6 day precedence in period 4. Such an
anti-correlation between the optical flux and polarization has been
found for other blazars with the Kanata telescope (Ikejiri et
al. 2011).  The optical polarization shows a correlation with the
gamma-ray flux in period 1 (Fig. \ref{fig:GvsPD}), and with the
optical color in periods 1, 2, and 4 (Fig. \ref{fig:VvsPD}, this is
known as a bluer-when-brighter trend).

We note several differences in the correlation trends between the four
periods.  Fig.\ref{fig:GvsV} shows that the optical flux is different
in different periods by a factor of 3--4 for the same gamma-ray flux.
If the origin of the gamma-ray emission is identical to that of the
optical emission and the environment of the emission region is the
same, the gamma-ray and optical fluxes should change together .

\begin{figure}[!htb]
\begin{center}
\FigureFile(80mm,50mm){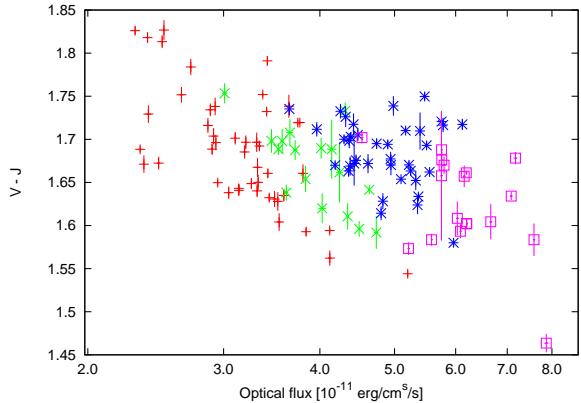}
\end{center}
\caption{Optical flux vs. color index ({\it V}-{\it J}). 
Colors are the same as those of Fig.\ref{fig:GvsV}.}
 \label{fig:VvsC}
\end{figure}

The PD is systematically different among the four periods
(Fig. \ref{fig:VvsPD}). This behavior is due to a slow change of the
PD in the long-term variable component. This trend is similar to that
found for 3C 454.3 (Sasada et al. 2010). An anti-correlation between
the optical flux and PD is seen during periods 3 and 4, and this could
be due to such a long-term variation.  
A bluer-when-brighter trend is seen within each period,  as well as
between periods 1 and 2 and between periods 3 and 4
(Fig. \ref{fig:VvsC}).  This is due to the long-term change of color
as reported for 3C 454.3 (Sasada et al. 2010).  However, the
color-flux relation is systematically different between periods 1--2
and 3--4; the color in periods 3--4 is redder than that in periods
1--2.  As a result, the correlation is weak over the full data set.
Assuming a new emission region appeared in periods 3--4, the
correlation would not be significant.  These observations imply at
least two components with different variability timescales in the
optical band.

\begin{table*}[!htb]
\caption{Correlation coefficients. The significance of the correlation
  was tested using the Student's t-test. Correlation coefficients whose
    significance is within the 95\% confidence interval are marked
    with an asterisk.}
\label{tab:correlation}
 \ \\
 \centering
   \begin{tabular}{lcccc}\hline\hline
& Period 1 & Period 2 & Period 3 & Period 4 \\ \hline
Gamma-ray vs {\it V} band & 0.53$\pm$0.22$^{*}$ & -0.10$\pm$0.16 & 0.19$\pm$0.84 & 0.02$\pm$0.18 \\
V-band vs PD & 0.35$\pm$0.18$^{*}$  & 0.05$\pm$0.25 & -0.34$\pm$0.20$^{*}$ & -0.26$\pm$0.14 \\
Gamma-ray vs PD & 0.82$\pm$0.20$^{*}$ & -0.13$\pm$0.22 & -0.90$\pm$0.58 &  0.11$\pm$0.17 \\
V-band vs Color ({\it V-J}) & -0.65$\pm$0.22$^{*}$ & -0.64$\pm$0.29$^{*}$ & -0.22$\pm$0.20 & -0.56$\pm$0.48$^{*}$ \\
    \hline
   \end{tabular}
\end{table*}

\section{Discussion} \label{sec:dis} 

Many FSRQs, such as PKS 1510-089 \citep{Abdo10f}, 3C 279
\citep{Abdo10e}, PKS 1502+106 \citep{Abdo10g}, and 3C 454.3
\citep{Striani10}, exhibit clear correlations between their gamma-ray
and optical properties.  Additionally, their broad-band spectral
energy distributions (SED) are better explained by SSC+EC models than 
SSC models \citep{Abdo10a}. These
prior measurements indicate that the radiation region of gamma-rays is
the same as that of the optical emission for FSRQs and LBLs.  From our
observations, 3C 66A presents two different behaviors: one with a
correlation between the gamma-ray and the optical bands, and another
uncorrelated one .  The correlation in period 1 in 2008 was also
reported in Paper I, and, for the first time during the \emph{Fermi}
era, the gamma-ray emission of this IBL can be explained by a SSC+EC
model (prior measurements of the IBL W Comae by VERITAS in TeV gamma
rays as part of a multiwavelength campaign suggested the emission was
SSC+EC based \cite{Acciari08}).  On the other hand, the optical flux
independently increases against the gamma-ray flux from period 1 to 4,
and no clear correlation is seen between the optical and gamma-ray
flux in periods 2, 3, and 4.  This type of behavior is novel for 3C
66A.

The average flux in the optical and gamma-ray bands in period 1 (2008)
and period 4 (2009) are listed in Table \ref{tab:ave_flux}.
\begin{table*}[!htb]
\caption{Average
    optical flux, optical color and gamma-ray flux.}
\label{tab:ave_flux}
 \ \\
 \centering
 \begin{tabular}{cccc} \hline\hline
  Period & Optical flux [erg cm$^{-2}$s$^{-1}$] & color ({\it V-J})&
  Gamma-ray flux at 20GeV [erg cm$^{-2}$s$^{-1}$] \\   \hline
  period 1 & (3.2 $\pm$ 0.9) $\times 10^{-11}$ & 1.68 $\pm$ 0.07 & (3.5 $\pm$ 1.6)$\times 10^{-11}$ \\
  period 4 & (6.0 $\pm$ 0.3) $\times 10^{-11}$ & 1.62 $\pm$ 0.06 & (3.8 $\pm$ 1.6)$\times 10^{-11}$ \\
  \hline
 \end{tabular}
 \\ \ \\
\end{table*}
The optical flux in period 4 is about twice as large as that in period 1 even
though the gamma-ray flux is similar .  In the framework of a one-zone
picture, we assume that the jet is oriented with respect to the line
of sight at a very small angle so that the Doppler factor is equal
to $D = (\Gamma [1-\beta \cos \theta_{\mathrm{obs}}])^{-1} = \Gamma$,
where $\theta_{\mathrm{obs}}$ is the angle of the jet with respect to
the line of sight and $\beta = (1-\Gamma^{-2})^{-1/2}$.

According to a simple SSC or SSC+EC scenario , if the flare is
assumed to be caused by a variation of the Lorentz factor
$\Gamma$(=$D$), the gamma-ray peak luminosity $L_{SSC}$ or $L_{EC}$
is predicted to vary as $L_{SSC} \propto D^{2+\gamma}$or $L_{EC}
\propto D^{2+2\gamma}$ where $\gamma$ is the photon index
\citep{Dermer95}.  In an SSC or SSC+EC model, if the flare is assumed
to be caused by electron input in the jet, the relation between
$L_{SSC}$ or $L_{EC}$ and $L_{sync}$ is expressed as, $L_{SSC}
\propto L_{sync}^2$ or $L_{EC}\propto L_{sync}$ where $L_{sync}$ is
the luminosity of the synchrotron emission of the jet.  In both cases,
the gamma-ray flux is expected to increase with the optical flux but
our results from period 4 do not match either scenario.  Since period
1 is explained by an EC model (Paper I), hereafter we only consider
the EC model.
 
The EC radiation luminosity ($L_{\mathrm{EC}}$) and synchrotron
radiation luminosity ($L_{\mathrm{sync}}$) are related by
\citep{Sikora08}:
\begin{equation}
\frac{L_{\mathrm{EC}}}{L_{\mathrm{sync}}} \sim \Gamma^2
\frac{u_{\mathrm{ext}}}{u_{B}}. \label{eq:sikora}
\end{equation}
Here,  $\Gamma$ is the bulk Lorentz factor of
the jet, $u_{\mathrm{ext}}$ is the energy density of
the external photons and $u_{\mathrm{B}}$ is the
energy density of the magnetic field.  Transforming
this expression with $u_{\mathrm{B}} = \frac{B^2}{8
  \pi}$ where $B$ is the amplitude of magnetic field, the value
of $u_{\mathrm{ext}}$ is then represented as
\begin{equation}
u_{\mathrm{ext}} \sim \frac{B ^2}{8\pi
\Gamma^2}\frac{L_{\mathrm{EC}}}{L_{\mathrm{sync}}}.
\label{eq:ext}
\end{equation}
The $L_{\mathrm{EC}}/L_{\mathrm{sync}}$ ratio depends on $\Gamma$,
$u_{\mathrm{ext}}$ and $u_B$.  As previously discussed, the variations
seen in this ratio are unlikely to be caused by variations in
$\Gamma$.  If variations of $u_B$ are the reason, then $B$ would have
to be stronger in period 4.  In general, the correlation between the
optical flux and PD should indicate whether the magnetic field is
strong, but our observations show an opposite trend.  Therefore
variations of $u_B$ are unlikely.  The
$L_{\mathrm{EC}}/L_{\mathrm{sync}}$ ratio also depends on the shape
of the electron energy distribution.  However, we find that the color
is almost the same between periods 1 and 4 (Table \ref{tab:ave_flux}),
and thus this possibility is unlikely.  Therefore, we focus on the
variations of the energy density of the external photon field
$u_{\mathrm{ext}}$ as the source of the variability.  Of course there
are other possible causes of the variability seen in the
$L_{\mathrm{EC}}/L_{\mathrm{sync}}$ ratio, such as the magnetic field
and the electron energy distribution varying at the same time.  In
this discussion, however, we focus on only one parameter for
simplicity.

We assume that the gamma-ray flux at 20 GeV, which is the peak energy
of EC radiation estimated from paper I, gives a peak luminosity of
the EC radiation, and also that both the optical and gamma-ray
emission originates from the same region.  We then calculate the
energy density of external photons for periods 1 and 4.  The
$u_{\mathrm{ext}}$ in period 1 is estimated in paper I, based on
fitting the multiwavelength spectra.  On the other hand, it is
difficult to estimate how much the EC radiation contributes to the
gamma-ray emission in period 4 using only the optical and gamma-ray
data.  Therefore, here we assume that the gamma-ray emission is
dominated by EC radiation.  We roughly calculate the energy density
of external photons and synchrotron photons, by using the magnetic
field $B$, the size of the emission region $R$, and the jet Lorentz
factor $\Gamma$ in the case of redshift z = 0.1, 0.2, 0.3, 0.444,
following Paper I.  The energy density of synchrotron photons is
represented as
\begin{equation}
 u_{\mathrm{sync}} = \frac{L{\mathrm{sync}}}{4\pi R^2 cD^4}.
\end{equation}

\begin{table*}[!htb]
\caption{Comparison of the estimated energy density of synchrotron and
  external photons inn 2008 (period 1) and 2009 (period 4) for each
  redshift.}
\label{tab:ratio}
 \ \\
 \centering
 {\small 
 \begin{tabular}{ccccc} \hline\hline
  Parameter & z = 0.1 & z = 0.2 & z = 0.3 & z = 0.444 \\   \hline
  \hline
  Comoving magnetic field,$B$[G] & 0.35 & 0.22 & 0.21 & 0.23 \\
  jet Lorentz factor $\Gamma$ & 30 & 30 & 40 & 50 \\
  size of blob $R$ [10$^{16}$cm] & 0.5 & 1.2 & 1.5 & 1.5 \\ \hline
  $u_{\mathrm{sync}}$  (period 1) [$10^{-5}$ erg/cm$^3$] & 10.56 & 8.36  &
  4.27  & 4.40 \\
  $u_{\mathrm{sync}}$  (period 4) [$10^{-5}$ erg/cm$^3$] & 19.60 & 15.51 & 7.92 & 8.17 \\
  $u_{\mathrm{ext}}$  (period 1) [$10^{-7}$ erg/cm$^3$]  & 58.61 & 22.98 & 11.78 & 9.04 \\
  $u_{\mathrm{ext}}$  (period 4) [$10^{-7}$ erg/cm$^3$]  & 29.21 & 11.54 & 5.92 & 4.54 \\
  \hline
 \end{tabular}
 \\ \ \\
 }
\end{table*}

In all cases, the $u_{\mathrm{ext}}$ in period 4 is about twice as
that in period 1.  $u_{\mathrm{ext}}$ can differ when the distance of
the gamma-ray emission region to the seed photon source is different
between periods 1 and 4.  For example, if the seed photons come from
the accretion disk, the distance to the central BH could be different
by a factor of only 1.4 between the two periods.  However, if the seed
photons come from the broad-line region or the dusty torus, the
situation is more complicated and beyond the scope of this work.  On
the other hand, the long-term correlations in the optical properties
indicate the existence of two emission components which could explain
the change of the $L_{\mathrm{EC}}/L_{\mathrm{sync}}$ ratio.

\citet{Marscher08} proposed a similar scenario that the radiation
source consists of two or more emission regions; one a global jet
region and the others a local one.  In this scenario, the local
emission is characterized by short-time variability (because of the
small emission region) and highly polarized emission (because the
small emission region aligns the magnetic field).  On the other hand,
the global emission consists of the addition of many local regions.
We conclude that the local emission is dominant in 2008, while the
global emission is dominant in 2009.

The brightening optical component in 2009 was associated with a bluer
color, but the PD decreased as the flux increased.  Such behavior is
also found in other blazars \citep{Ikejiri11}, and could be explained
by the existence of an underlying constant component, and a short-term
variable component with a different polarization direction
\citep{Uemura10}. In this case, the underlying component has a
polarization with PD$\sim$13 [\%] and PA$\sim$200$\arcdeg$, while the variable
component that causes the flux variation in periods 1 and 4 has a
polarization whose direction is quite different.  A gradual shift of
PA in periods 1 and 2 supports the two-component model.

\section{Summary}
We performed long-term monitoring of the ISP blazar 3C 66A with the
LAT onboard the \emph{Fermi Gamma-ray Space Telescope} and the Kanata
telescope and studied the correlations among various gamma-ray and
optical properties.  As a result, we find two distinct types of
behavior: one, in 2008, which shows a correlation between the optical
properties and the gamma-ray flux and the other , in 2009, which does
not show a good correlation.  This result indicates that the emission
region is different between these periods.  Paper I indicates that the
emission from this source is well explained by a SSC + EC model
during the 2008 flare and we estimated the environment of the jet in 2009
using same model mentioned in Paper I.  Based on this assumption, we
calculated the $u_{\mathrm{ext}}$ value in each state and found that
it is different by a factor of two.  Those different behaviors between
the gamma-ray and optical bands might be explained by postulating two
different emission components.

\bigskip

\section{Acknowledgments}

The \textit{Fermi} LAT Collaboration acknowledges generous ongoing support
from a number of agencies and institutes that have supported both the
development and the operation of the LAT as well as scientific data analysis.
These include the National Aeronautics and Space Administration and the
Department of Energy in the United States, the Commissariat \`a l'Energie Atomique
and the Centre National de la Recherche Scientifique / Institut National de Physique
Nucl\'eaire et de Physique des Particules in France, the Agenzia Spaziale Italiana
and the Istituto Nazionale di Fisica Nucleare in Italy, the Ministry of Education,
Culture, Sports, Science and Technology (MEXT), High Energy Accelerator Research
Organization (KEK) and Japan Aerospace Exploration Agency (JAXA) in Japan, and
the K.~A.~Wallenberg Foundation, the Swedish Research Council and the
Swedish National Space Board in Sweden.

Additional support for science analysis during the operations phase is gratefully
acknowledged from the Istituto Nazionale di Astrofisica in Italy and the Centre 
National d'\'Etudes Spatiales in France.

This work was supported by Japan Society for the Promotion of Science(JSPS).


\begin{thebibliography}{}
\bibitem[{Abdo} {et~al.}(2010a)]{Abdo10a}                                
Abdo,~A.~A., et~al. 2010a, \apj, 715, 429
\bibitem[{Abdo} {et~al.}(2010b)]{Abdo10c}                                
Abdo,~A.~A., et~al. 2010b, \apjs, 187, 460
\bibitem[{Abdo} {et~al.}(2010c)]{Abdo10e}                                
Abdo,~A.~A., et~al. 2010c, \nat, 463, 919
\bibitem[{Abdo} {et~al.}(2010d)]{Abdo10Cat}                                
Abdo,~A.~A., et~al. 2010d, \apjs, 188, 405
\bibitem[{Abdo} {et~al.}(2010e)]{Abdo10f}                                
Abdo,~A.~A., et~al. 2010e, \apj, 721, 1425
\bibitem[{Abdo} {et~al.}(2010f)]{Abdo10g}                                
Abdo,~A.~A., et~al. 2010f, \apj, 710, 810
\bibitem[{Abdo} {et~al.}(2010g)]{Abdo10i}                                
Abdo,~A.~A., et~al. 2010h, \apj, 710, 1271
\bibitem[{Abdo} {et~al.}(2011)]{Abdo10d}                 
Abdo,~A.~A., et~al. 2011, \apj, 726, 43
\bibitem[{Acciari} {et~al.}(2008)]{Acciari08}
Acciari,~V.~A., et~al. 2008, \apj L, 684, 73 
\bibitem[{Atwood} {et~al.}(2009)]{Atwood09}
Atwood, W. B., et~al. 2009, \apj, 697, 1071
\bibitem[{Bramel} {et~al.}(2005)]{Bramel05}
Bramel, D. A., et~al. 2005, \apj, 629, 108
\bibitem[{Cutri} {et~al.}(2003)]{Cutri03}
Cutri., et~al. 2003, 2MASS All Sky Catalog of Point Sources, The IRSA 2MASS All-Sky Point Source Catalog, NASA/IPAC Infrared Science Archive. \\
http://irsa.ipac.caltech.edu/applications/Gator/ 
\bibitem[{Dermer} {et~al.}(1995)]{Dermer95}
Dermer C. et~al. 1995, \apj, 446, 63
\bibitem[{Edelson \& Krolik}(1998)]{Edelson98}
Edelson, R. A. \& Krolik, J. H. 1998, \apj, 333, 646
\bibitem[{Gonzalez-Perez} {et~al.}(2001)]{Gonzalez01}
Gonzalez-Perez, J. N., et~al. 2001, \apj, 122, 2055
\bibitem[{Ikejiri} {et~al.}(2011)]{Ikejiri11}
Ikejiri, Y., et~al. 2011, \pasj, 63, 639
\bibitem[{Marscher} {et~al.}(2008)]{Marscher08}
Marscher, A. P., et~al. 2008, \nat, 452, 966
\bibitem[{Miller} {et~al.}(1978)]{Miller78}
Miller, J. S., French, H. B., \& Hawley, S. A. 1978, Proc. Pittsburgh Conf. on {\it BL Lac Objects}, ed. A. M. Wolfe,
\bibitem[{Nolan} {et~al.}(2012)]{Abdo12Cat}                              
Nolan,~P.~L., et~al. 2012, \apjs, 199, 31
\bibitem[{Sasada} {et~al.}(2010)]{Sasada10}
Sasada, M., et~al. 2010, \pasj, 62, 645
\bibitem[{Schlegel} {et~al.}(1998)]{Schlegel98}
Schlegel, D., et~al. 1998, \apj, 500, 525
\bibitem[{Sikora} {et~al.}(2008)]{Sikora08}
Sikora, M., et~al. 2008, \apj, 675, 71
\bibitem[{Striani} {et~al.}(2010)]{Striani10}
Striani,~E. et~al. 2010, \apj, 718, 455
\bibitem[{Uemura} {et~al.}(2010)]{Uemura10}
Uemura, M., et~al. 2010, \pasj, 62, 69
\bibitem[{Watanabe} {et~al.}(2005)]{Watanabe05}
Watanabe, M., et~al. 2005, \pasp, 117, 870
\bibitem[{Wolff} {et~al.}(1996)]{Wolff96}
Wolff, M. J., et~al. 1996, \aj, 111, 856
\bibitem[{Wright, E. L.} (2006)]{Wright06}
Wright, E. L. 2006, \pasp, 118, 1711
\end{thebibliography}
\end{document}